# Room temperature large spontaneous exchange bias in hard-soft antiferromagnetic composite $BiFeO_3$-$TbMnO_3$

Prince Kr. Gupta[1], Surajit Ghosh[1], Shiv Kumar[2], Arkadeb Pal[1], Rahul Singh[1], Abhishek Singh[1], Somnath Roy3, Eike F. Schwier[2] , Masahiro Sawada[2], Takeshi Matsumura[2], Kenya Shimada[3], Hong-Ji Lin[4], Yi-Ying Chin[5], A.K. Ghosh[3] and Sandip Chatterjee[1,]*

[1]Department of Physics, Indian Institute of Technology (Banaras Hindu University), Varanasi-221005, U.P. India

[2] Hiroshima Synchrotron Radiation Center, Hiroshima University, Kagamiyama 2-313, Higashi-Hiroshima 739-0046, Japan

[3] Department of Physics, Banaras Hindu University, Varanasi – 221005, U.P. India

[4]National Synchrotron Radiation Research Center, Hsinchu 300, Taiwan

[5]Department of Physics, National Chung Cheng University, 168, Sec. 1, University Road, Min-Hsiung, Chiayi 62102, Taiwan

*Corresponding author e-mail id: schatterji.app@iitbhu.ac.in

## Abstract

We report the presence of giant spontaneous exchange bias ($H_{SEB}$) in a hard and soft antiferromagnetic composite of $BiFeO_3$-$TbMnO_3$ (BFO-TMO in 7:3 and 8:2 ratio). The $H_{SEB}$ varies between 5-778Oe, but persists up to room temperature with a maximum near a spin reorientation transition temperature observed from magnetization vs. temperature measurement in Zero-field cooled (ZFC) and Field cooled (FC) modes. Isothermal remnant magnetization measurements at room temperature indicate the presence of an interfacial layer of a 2 dimensional dilute antiferromagnet in a field (2D DAFF). A stable value of the exchange bias has been observed via training effect measurements which signify the role of interfacial exchange coupling in the system. Based on the experimental results we explain the presence of the giant spontaneous exchange bias on the basis of a strong strain-mediated magnetoelectric coupling induced exchange interaction and the creation of 2D DAFF layer at the interface. The properties of this layer are defined by canting and pinning of BFO spins at the interface with TMO due to Fe and Mn interaction. X-ray Magnetic Circular Dichroism (XMCD) confirms the presence of canted antiferromagnetic ordering of $BiFeO_3$, charge transfer between Mn ions and different magnetically coupled layers which play vital role in getting the exchange bias.

## 1. Introduction

Since its discovery by Meiklejohn and Bean in 1956 the Exchange bias (EB) phenomenon has been the focus of extensive theoretical and experimental analysis and more recently it is attracting researchers for its potential application in devices like giant magnetoresistance, high-density data storage, electrically controllable spin-polarized currents, magnetic sensors and spin valve devices [1-5]. Conventional exchange bias (CEB) is observed at the interface between ferromagnetic and antiferromagnetic materials when cooled in a field through the Neel temperature ($T_N$) of the hard magnetic material [1, 6-7]. The spontaneous exchange bias (SEB) on the other hand has gained much attention as the unidirectional anisotropy develops spontaneously, even if the material is not cooled in a field through $T_N$. Apart from interacting interfaces between FM and AFM materials, the effect has so far been observed in different types of heterostructures, such as, ferromagnetic–antiferromagnetic, ferromagnetic–spin glass–antiferromagnetic, ferromagnetic–ferrimagnetic, ferromagnetic–dilute antiferromagnetic etc. in different geometries like bilayers, core-shell nanostructures, granular composites and superlattices [2, 8-10]. Overall the CEB is found more commonly in the literature, whereas the SEB has been reported for a smaller number of materials such as $BiFeO_3$-$Bi_2Fe_4O_9$ nanocomposite, $Mn_2PtGa$ Heusler alloys and polycrystalline $Co_{0.75}Cu_{0.25}Cr_2O_4$ samples [11-13].

In recent years $BiFeO_3$ has emerged as one of the most significant materials for investigating the EB as well as multiferroicity. It is the only perovskite material which shows multiferroic behavior at room temperature (Neel temperature $T_N \approx 643$ K and Curie temperature $T_C \approx 1100$ K) [14]. Different low dimensional $BiFeO_3$ systems, such as nanoparticles and nanotubes are known to display weak ferromagnetic behaviour at room temperature and spin glass type behaviour at low temperature [15-17]. The EB and its origin in $BiFeO_3$ based systems are in discussion in recent times [11, 15, 18-23]. It was found to persist at room temperature in some nano-dimensional systems consisting $BiFeO_3$ and was ascribed to the spin pinning at the interface between the AFM and weak FM layers present in those structures [15]. Dong *et al*. have demonstrated that SEB exists in $BiFeO_3$ nanocrystals with a non-monotonic temperature dependence of $H_{SEB}$. They proposed that the formation of a diluted antiferromagnetic phase and its interaction with the hard AFM moments was the mechanism responsible for the high-temperature behavior of $H_{SEB}$. The low-temperature $H_{SEB}$ were explained by suggesting the formation of super spin-glass (SSG) type structure and related interactions [18]. Manna *et al.* have reported $H_{EB}$ above the SSG transition within their $BiFe_{0.8}Mn_{0.2}O_3$ nanoparticles and have explained the EB on the basis of interaction between antiferromagnetic core and a diluted antiferromagnetic (DAFF) shell [19]. Zhang *et al*. have also shown room temperature exchange bias in $BiFeO_3$ nanocrystals and explained the phenomenon as a consequence of interaction between antiferromagnetic core and DAFF shell. [23]. Recently, Maity *et al*. on the other hand found EB only when a SSG-type structure is formed at low temperature in $BiFeO_3$-$Bi_2Fe_4O_9$ nanostructures and the phenomenon was linked to pinning of glassy spins to the remaining antiferromagnetic spins [11, 21]. A skin layer different from its bulk is also reported in single crystalline $BiFeO_3$ which influences several of its properties including the EB [24].

Several attempts have been made to induce chemical pressure in the $BiFeO_3$ lattice by substituting Bi or Fe ions with rare earth and transition metal atoms respectively. Substitution is known to modify the magnetic behaviour and enhance the EB [22]. Attempts have also been made to integrate $BiFeO_3$ in heterostructures and composite systems containing other perovskite materials like, $BaTiO_3$and $SrTiO_3$ [25, 26]. However, a clear understanding of the mechanism which leads to the EB effect is still missing. Although, the SEB effect has been investigated for many years, obtaining a large $H_{SEB}$ at room temperature and within the bulk state remains an ongoing task since room temperature EB is more attractive for device applications.

In this context, we decided to incorporate $TbMnO_3$ (which shows very strong magneto-electric coupling) into $BiFeO_3$. $TbMnO_3$ is a promising material to induce strain at the single phase interfaces in the system [27]. $TbMnO_3$ possesses a perovskite structure (space group *Pbnm*) and the lattice mismatch between the two materials in the composite system is sufficient to create strain [27]. In this article we examine composites of 0.7BFO-0.3TMO (70% $BiFeO_3$ and 30% $TbMnO_3$) and 0.8BFO-0.2TMO (80% $BiFeO_3$ and 20% $TbMnO_3$). We found (i) a large value of spontaneous exchange bias both at room temperature and low temperatures (ii) a non-monotonic variation of exchange bias with temperature. We propose an exchange interaction initiated by the strain-mediated strong magnetoelectric coupling which is the result of lattice mismatch between the two materials. Further, we demonstrated by means of XMCD that the canted nature of antiferromagnetic ordering in BFO, exchange interaction between Fe and Mn ions present in the system and the pinning layer at the interface play a vital role in determining the amplitude and temperature behavior of the EB.

## 2. Experimental methods

The synthesis of magnetoelectric multiferroics composite consisting of $BiFeO_3$ in the bulk via solid state reaction is a difficult task because it commonly leads to the formation of thermodynamically stable impurities $Bi_2Fe_4O_9$, $Bi_{25}FeO_{39}$ and $Bi_{46}Fe_2O_{72}$ due to the volatile nature of $Bi_2O_3$ [28]. In this work, we have used high purity oxides $Bi_2O_3$, $Fe_2O_3$, $Mn_2O_3$, and $Tb_4O_7$ as starting materials. First, polycrystalline $TbMnO_3$ was prepared through a solid state reaction as described in our previous report [29]. Then stoichiometric amounts of the materials $Bi_2O_3$ (5% excess), $Fe_2O_3$ and $TbMnO_3$ were mixed for 3-4 hours and then the mixture was calcined at 1173 K for 1 hour to prepare composites with two different ratios (70:30 and 80:20) namely, 0.7BFO-0.3TMO and 0.8BFO-0.2TMO. The calcined powder was ground and pressed into pellets and sintered for 2 hours at 1273 K with some mixture of calcined powder as a spacer to reduce the loss of $Bi_2O_3$ during synthesis.

The structural analysis of the composite was carried out using an X-ray diffractometer (Model: Miniflex-II, Rigaku, Japan) with Cu Kα radiation ($\lambda$ = 1.5406 Å) in step size of 0.002 with a scan speed of 2°/min. For their magnetic properties, the composites were investigated using a commercial superconducting quantum interference device [Magnetic Properties Measurement System XL-7, Quantum Design, Inc.] magnetometer. [X-ray absorption spectroscopy (XAS) and x-ray magnetic circular dichroism (XMCD) measurements at Mn $L_{2,3}$-edge and Fe $L_{2,3}$- edge at 300 K and 180 K were performed with polarized X-rays at beamline from Hsinchu Synchroton center, Taiwan.

## 3. Results and discussion

### A. X-ray analysis

Figure 1 shows the X-ray diffraction profile for the composite 0.7BFO-0.3TMO in the 2θ range from 20° to 80°. All the observed peaks in X-ray profile correspond to individual phases of BFO and TMO and matched well with the standard JCPDS data (Card No. 74 2016 for $BiFeO_3$ and Card No. 250 933 for $TbMnO_3$). No other peaks corresponding to any intermediate phase and/or impurity could be found indicating the existence of pure BFO and TMO phases without any chemical reaction between them. A closer look at the 2θ value 32° which corresponds to 110 plane of BFO in *R3c* phase reveals that the peak is shifted to the higher angle side as well as split into multiple peaks which hints towards the existance of multiple phases of BFO. Rietveld refinement using the FULLPROF package was employed to index the XRD pattern for analysing the crystal structure and structural phase transition from the intensities of overlapping reflections in the XRD pattern. The difference between experimental and theoretical pattern as well as Bragg positions are shown in the bottom of the plot. From the refinement result it was found that in the composite, $TbMnO_3$ crystallises as in its native bulk structure i.e.,orthorhombically distorted perovskite structure with space group *Pbnm.* $BiFeO_3$ on the other hand is found to co-exists in two very similar phases consisting of rhombohedral and orthorhombic symmetry (*R3c*+*Pbnm*).

Rare-earth orthoferrites $RFeO_3$($R$= rare earth), manganites and the high-temperature phase of BFO are known to undergo a structural transition from the rhombohedral *R3c* structure to an orthorhombic (*Pnma or Pbnm*) structure when sintered at higher temperature [30]. The observation of the existence of dual phases i.e., rhombohedral (*R3c*) and orthorhombic (*Pnma or Pbnm*) in rare earth (RE)-modified BFO has previously been reported **[**31**]**. Recent reports show that BFO in solid solution with rare earth manganites ($RMnO_3$) i.e., $BiFeO_3$-$RMnO_3$ ($R$=La, Gd, Ho and Dy), and (1-x)$BiFeO_3$-x$YMnO_3$ undergo a structural phase transition from *R3c* to *R3c* + *Pnma* over a wide compositional range [32-34]. This structural transition to dual phases was accompanied by a significant improvement of multiferroic properties within BFO. Lotey *et al*. reported complete structural transformation from rhombohedral (*R3c*) to the orthorhombic ($Pn2_1a$) phase at 15% of Tb-doping in BFO nano-particles [35]. The existence of both rhombohedral and orthorhombic phase in the composite can therefore be attributed to (a) the deficiency of $6s^2$ lone pair in $Tb^{+3}$ hinders the stereochemical activity of the lone pair of $Bi^{+3}$ ions in the structure (b) $Mn^{+3}$ weakly destabilizes *R3c* phase whereas $Fe^{3+}$ stabilizes it (c) chemical pressure of $Tb^{3+}$ is much smaller than for $Bi^{3+}$ as the ionic radii of $Bi^{3+}$ and $Tb^{3+}$ are 1.31 A˚ and 1.17 A˚, respectively, for high-spin eight-fold coordination [36]. The two phases in our sample correspond to bulk (*R3c*) and interfacial (*Pbnm*) phase at the interface with $TbMnO_3$. The interfacial regions of the $BiFeO_3$ crystals face strain due to the presence of $Tb^{3+}$ and $Mn^{3+}$ and the interface strucutre changes to the *Pbnm* phase.

## B. Magnetic Properties

Figure 2 presents the M-H loop at different temperatures of the composite 0.7BFO-0.3TMO. The M-H loop at room temperature shows a linear variation of magnetization with an applied magnetic field which is typical for an antiferromagnetic material. Around the zero field, a deviation of non-linearity is observed which indicates the existence of FM ordering too. Furthermore, a close view around the zero field (inset of figure 2) reveals that the loops are asymmetric in nature, yielding a spontaneous exchange bias (SEB) for 0.7BFO-0.3TMO at all the temperatures. The presence of large loop shift in the room temperature M-H loop can be quantified as exchange bias field, $H_{SEB}= [(H_{C1}+H_{C2})/2])$, where $H_{C1}$ and $H_{C2}$ are the points in the field axis where magnetization value becomes zero. The coercivity has also been estimated from the M-H loop as $H_C = [(H_{C1}-H_{C2})/2]$. The value of $H_{SEB}$ at room temperature was ~ 510 Oe which is very high in comparison with other BFO based systems which show room temperature EB. From the plots it can be found that the loops are also shifted, hence producing the EB field for 0.8BFO-0.2TMO composite as well. $H_{SEB}$ estimated at different temperature are plotted in figure 3 for both composites. From the plot, it becomes evident that $H_{SEB}$ exhibits clear temperature dependence in both composites with similar nature. Both the composites show initial increase in the exchange bias on decreasing the temperature. $H_{SEB}$ values show a decreasing trend on further decreasing the temperature down to 5 K. 0.7BFO-0.3TMO shows maximum $H_{SEB}$ at 225 K whereas 0.8BFO-0.2TMO shows maximum EB at 180 K. The inset of the figure 2 shows M-H plot measured following two different measurement protocols (P-type and N-type), with different starting fields which have opposite signs. It can be seen from the plot that the M-H loops are shifted to the left of the origin independent of the starting field which indicates the presence of pinned magnetic moments in the system. In general for spontaneous exchange bias without the presence of pinned magnetic moments show symmetric M-H loops (i.e., almost equally shifted from the origin on either side depending on the measurement protocol) [4].

Figure 4a and b depicts the temperature dependence of field cooling (FC) and zero field cooling (ZFC) dc-magnetization for composite 0.7BFO-0.3TMO and 0.8BFO-0.2TMO, respectively. The two different compositions show similar behaviors in the ZFC-FC magnetization where the ZFC and FC values increases with decreasing temperature at low temperatures. In this plot (figure 4a), the magnetization (both FC and ZFC) for the composition 0.7BFO-0.3TMO is observed to decrease with decreasing temperature in the temperature range 215 – 300 K which is expected for an antiferromagnetic material below its Neel temperature. Further, on decreasing the temperature below 215 K (T*), FC and ZFC magnetization curves show upturns. To get clear view of the anomaly the dM/dT vs T is plotted in the insets of the figure 5a and b from where it is found that in case of 0.7BFO-0.3TMO the slope of the dM/dT vs T curves changes abruptly at ~ 215 K whereas for the other composite the transition temperature was found to be 169 K. Similar trends have previously been reported as a result of reorientations of spins with the involvement of electromagnons, similar to that observed in orthoferrites [37-42]. The transition is reported to be of magnetic origin and not related to structural deformation or transition [42]. Although the spin reorientation transition is reported well above the room temperature and just below the $T_N$ for different composite systems such as, $BiFeO_3$-$PbTiO_3$ and $BiFeO_3$-$BaTiO_3$, T* has been reported in the temperature range of 150-200 K for pure $BiFeO_3$ [41, 42]. Recently, A. Kumar *et al.* have also observed such a spin reorientations transition in pure and 0.3% $MnO_2$ doped $BiFeO_3$ powders in the same temperature range [42]. Interestingly, the maximum EB was observed at ~ 180 K and ~ 225 K which are just above the spin reorientation transition temperatures (169 K and 215 K) for 0.8BFO-0.2TMO and 0.7BFO-0.3TMO respectively. Below the transition as the temperature decreases the $H_{SEB}$ value decreases. In several systems, it has been observed that the exchange bias gets modified, (even sign of exchange bias gets reversed), around the spin reorientation transition in presence of Fe ions [43, 44]. In few systems also similar trend of getting the maximum exchange bias near the spin reorientation has been observed and was explained on the basis of compensation of the interfacial spins below the spin reorientation transition [43]. Further decreasing the temperature, FC and ZFC magnetization curves show very low bifurcation from 95 K to 2 K as the two curves deviate from each other very slowly towards lower temperatures. Although there is bifurcation in the plots, we have not found any traces of long range or short range ordering from either the dM/dT vs T plots or AC susceptibility measurements (not shown here) in both the sample. $TbMnO_3$ is known to exhibit AFM ordering below 40 K in the bulk, but feature is also not traceable in M-T plots [29].

The training effect is one of the important properties of EB system as it is a measure of the stability of the exchange bias effect against thermal cycling. The effect is manifested as a gradual reduction in EB when subsequent recording of the hysteresis loop is performed several times. In order to understand the training effect, consecutive hysteresis loops (n=1-18) were recorded at room temperature for the composite systems in ZFC mode (shown in figure 5 a and b). Further, the evaluation of $H_{SEB}$ with n confirmed the presence of the training effect in our samples. It can be clearly seen from the plot that the M-H loop changes abruptly in between first and second loops and becomes stable very soon after second hysteresis loop. This variation can be understood by training effect model related to symmetry of antiferromagnet given by Hoffmann [45]. According to him the non-collinear spin structure changes to collinear spin structure between the first and second loop due to the presence of multiple antiferromagnetic easy anisotropy axes. In general, the training effect can be understood as a result of the relaxation of the uncompensated spins at the interfacial region between the antiferromagnetic and ferromagnetic layer. These rearranged spins are contributing to the exchange bias effect when exposed to a repeating measurement of the hysteresis loop along with thermal cycling [46-57]. The inset of figure 5 a and b shows the variation of the exchange bias plotted against the number of loops. As can be seen from the plot, the value of $H_{SEB}$ decreases sharply after the first cycle and then tends to saturate after a few cycles for both the samples. This small change in training effect for higher loops can be quantified by the following empirical power law;

$$H_{SEB}(n) - H_{SEB}(\infty) \propto \frac{1}{n^{1/2}} \qquad (1)$$

where n is the number of the loops and, $H_{SEB}$ (∞) is the spontaneous EB field in the limit of infinite loops. The solid red line in figure 5 represents the best fitting result of eq. 1 with field cycle number n≥2. The obtained value of $H_{SEB}$ (∞) from the fit is very large at room temperature and close to the value observed in the first loop. The obtained fitting values of $H_{SEB}$ (∞) = 596 Oe for 0.7BFO-0.3TMO and 1100 Oe for 0.8BFO-0.2TMO gives the satisfactory result with experimental data. Although eq. 1 gives satisfactory fitting at higher value of n, but it fails to give proper fitting at the data point at n=1. To describe the variation $H_{SEB}$ with n at higher as well as lower value of n, a more generalized recursive equation (given below) proposed by Binek *et al.* [47], can be used;

$$[H_{SEB}(n+1) - H_{SEB}(n)] = -\gamma[\{H_{SEB}(n) - H_{SEB}(\infty)\}]^3 \tag{2}$$

where $\gamma$ is a sample-dependent constant. The solid red line in the inset of figure 5a and 5b represents the fitting of the training data points according to eq. 1 and 2 respectively. It can be seen that for both the samples eq. 1 fits well with higher loop values $n \geq 2$ whereas eq. 2 fits well for data points starting from n=1. A mere change of ~ 3 to 4% has been observed after 18 loops for the samples 0.7BFO-0.3TMO and 0.8BFO-0.2TMO respectively, which implies that the uncompensated spins are quite stable under thermal cycling and that the SEB is robust. This exponential decrease in $H_{SEB}$ implies that there are few unstable uncompensated spins at the interfacial region of the multiple-domain state of 0.7BFO-0.3TMO the energy of which got reduced after cycling through the consecutive hysteresis [4, 47-48]. The training effect has been shown in different systems showing EB, such as, double perovskite compounds, FM-AFM bilayers, FM nanodomains embedded in an AFM matrix, core-shell nanoparticles etc. In those cases, the EB has been shown to originate only from the interfacial exchange coupling. Binek's model of training effect has been successfully applied to AFM core and 2D DAFF shell where the interfacial exchange coupling between the core and the shell also plays a vital role comparable to the BFO-TMO system investigated in this paper [19, 47].

Recently, isothermal remanent magnetization (IRM) measurement is being used as a finger printing technique to differentiate different magnetic states such as DAFF, spin glass and others as it identifies the nature of the irreversible magnetization contributions [49]. We have performed IRM vs field measurements to look out for the presence of any interface pinning containing hard antiferromagnet $BiFeO_3$ and possible dilute antiferromagnet. To measure the IRM, a field is applied on the sample for a very short time (~ 60 s) after cooling it from a high temperature and then the remanent magnetization is immediately recorded. It is expected that a pure AFM material shows zero IRM value for all fields and all temperatures as the reversible magnetization becomes zero for a pure AF state. From the figure 6 it can be seen that the IRM value observed is very low and increases little when the field is raised to 1.5 T where the virgin curve increases sharply and linearly with the field. This behavior is consistent with an AFM state. However, the IRM is not showing a typical behavior of a 3D dilute antiferromagnet as it is not expected to increase at all. This type of behavior of IRM has been found for $Co_3O_4$ nanowires where the nature of the particle is found to be of 2D DAFF [50]. The 2D DAFF layer is attributed to the formation of a 2$^{nd}$ phase (*Pbnm*) of BFO at the interfacial regions other than that of core BFO (*R3c*). The antiferromagnetic ordering in the two phases is expected to be different [24]. However, the origin of the DAFF layer remains in doubt as it could be formed either due to BFO skin layer or due to the TMO spins which could be coupled to the interfacial spins of the BFO.

The BFO/TMO interface is likely to be responsible for the EB and is expected to exhibit a complex interface structure, including charge transfer, atomic spin and orbital configurations. At the interface of the composite two different phases are combined at the atomic level which results in an increased magnetoelectric (ME) coupling due to the strain-mediated ME effects across the interfaces [45]. This enhanced ME coupling can modify the lattice structure through spin reconstruction at the interface as seen from the XRD results. From the calculation of magnetic coupling across the interface of $BiFeO_3$ and other manganites, it has been found that both charge and orbital ordering at the interface result in developing a magnetic moment of Fe ions at the BFO-TMO interface [20]. Another possibility can be the presence of an orbital reconstruction which will lead to a strong hybridization between Fe $d_{3z2-r2}$ orbital and Mn $d_{x2-y2}$ via an oxygen mediated superexchange [48]. According to the Anderson-Goodenough-Kanamori rules, the exchange interaction between Fe and Mn cations are expected to be ferromagnetic and the Mn-O-Mn interaction to be antiferromagnetic [51]. Thus, there lies a competition between the interfacial interaction of the Fe and Mn spins and the bulk antiferromagnetic interaction of the BFO which results into the canting of the interfacial spins. The exchange bias can be understood due to the pining and canting of the interfacial spins to the core antiferromagnetic spins to BFO. Moreover, the possibility of charge flow across the interface can be neglected as there is a large difference between the energy level of BFO and TMO, as estimated from atomic stacking $[(BiO)^{+}$-$(FeO_2)^{-}$-$(BiO)^{+}$-$(MnO_2)$-$(TbO)^{+}]$ equation for BFO/TMO [20, 52]. The strain mediated ME coupling also produces a very thin layer of BFO with the different structure as seen from the XRD results which can be the reason behind getting the 2D DAFF layer at the interface observed from the IRM measurement (figure 6) which also plays a significant role in getting the exchange bias. Thus, the large coercivity can also be explained on the basis of interfacial ferromagnetic superexchange interaction. At room temperature, the value of exchange bias got reduced from the maximum value at 225 K due to the large thermal fluctuations of spins resulting into less interfacial coupling [18].

The origin of exchange bias in $BiFeO_3$ based systems has been an argued mechanism in recent years. As mentioned during the introduction, some $BiFeO_3$ based systems exhibit EB only below super spin glass (SSG) transition temperatures where the explanation was the presence of SSG moments [11, 21]. The SSG moments at the core generate a random field that can induce a variation in the anisotropy of the AFM moments including biaxiality with respect to the direction of the applied field and set the uniaxial anisotropy via RKKY interaction [21]. Our ZFC-FC M-T data shows little bifurcation below ~95 K for 0.7BFO-0.3TMO suggesting weak frustration in the system, but in case of 0.8BFO-0.2TMO there is no bifurcation in the ZFC-FC data. We also could not find signals of any relaxation phenomena from our AC susceptibility measurements for both the samples (not shown here). Therefore, we could rule out the possibility of any kind of SSG mediated phenomena happening to give rise to the EB effect. Thus, the results obtained from our study cannot be explained within the model given by Maity *et al*. [11, 21]. Manna *et al*. found room temperature and low temperature exchange bias in their Mn-doped $BiFeO_3$ nanoparticles [19]. The results were understood on the basis of a core-shell model of hard antiferromagnetic core and dilute antiferromagnet shell where the spins of DAFF gets pinned to the core's spins when a magnetic field is applied, thus, setting the uniaxial anisotropy in the system. Zhang *et al.* also found DAFF layer to influence the magnetic property of the $BiFeO_3$ nanoparticles [23]. Therefore, to verify the compatibility of this model with our system a confirmation of type of interaction between the different ions present in the composite in different temperature is necessary. To this end we have recorded X-ray absorption (XAS) and X-ray magnetic circular dichroism (XMCD) spectra of Fe and Mn $L_{2,3}$ edge.

### C. XAS and XMCD Results

The synchrotron based XAS is a spectroscopic technique which probes electronic states of a matter. In x-ray absorption spectroscopy (XAS), x-ray is made incident on the core level electrons to excite them to unoccupied valence levels. The transition from core to valence level is governed by the dipole selection rules. Thus XAS has direct correspondence with unoccupied density of states (DOS). As it has a direct correlation with the unoccupied density of states (DOS) several information regarding oxidation states, local symmetries, and the spin and orbital magnetic moments of the Fe and Mn ions and interaction between them in the composite can be investigated by means of XAS and XMCD measurements.

The XMCD signal at the Mn and Fe $L_{2,3}$ edges (excitation from filled 2p→3d transition) is obtained from the difference between the two XAS spectra, taken as the difference between XAS spectra recorded with left and right hand circularly polarized light ( $\Delta\mu = \mu^+ - \mu^-$) in presence of a magnetic field of 0.6T. XMCD measurements were performed in total electron yield (TEY) mode at 180 K and 300 K. Figures 7a-d depicts the XAS and XMCD spectra of Fe $L_{2,3}$ of the two composites measured at 300 K and 180 K. The spectral shape and energy position in the Fe $L_{2,3}$-edge splits due to spin-orbit coupling at $L_3(2P_{3/2})$ at ~710eV and $L_2(2P_{1/2})$ at ~ 722eV corresponding to the absorption edge of $Fe^{3+}$ions. The $L_3$ and $L_2$ peaks are also split due to the crystal field to doubly degenerate $e_g$ and triply degenerate $t_{2g}$ levels [53]. The shape of the XAS peak of BFO matches well with the $Fe^{3+}$ signal previously seen in $BiFeO_3$ based systems [54, 55]. The shape of the spectral lines for the composites matches with the calculated spectrum taking the high spin configuration of $Fe^{3+}$ ions ($t_{2g}^3 e_g^2$) and with the measured $Fe^{3+}$ XAS signal [56]. A closer look at the $L_3$ edge of the composites (inset of figure 7a) when compared with the Fe $L_3$ edge of standard α-$Fe_2O_3$ reveals that the spectral shape of the $t_{2g}$ peak of the $L_3$ edge is diminished and shifted to the higher energy side for the composites. The overall spectra of the composites are also broader in comparison to α-$Fe_2O_3$. These features of the $t_{2g}$ can arise from the presence of either different crystalline coordination of the Fe ions or of different valence states of Fe in the sample [57, 58]. The $L_2$ edge of the composites shows $Fe_3O_4$ like diminished $L_2$ edge from where one might get the indication of mixed valence state of Fe. But, a closer inspection of the $L_2$ edge (inset of figure 7b) supports the fact that Fe ions are in mixed crystal coordination as the peaks are not shifted to the lower energy side [59] which should have been the case for mixed valence state. From the XAS study, Fe ions are thus, found to be in trivalent state with octahedral and another crystalline coordination.

The bottom panels of figure 7 (a-d) show the XMCD spectra of Fe ions of the composite 0.7BFO-0.3TMO. The XMCD signal in the composite system is weak and in the figure 7 the XMCD signals are magnified five times for clear visualization. $BiFeO_3$ is antiferromagnetic at room temperature and hence it is not dichroic thus, the absence of the XMCD signal of the composites at room temperature can be understood. Interestingly, The Fe XMCD spectrum at 180 K for the 0.8BFO-0.2TMO shows clear dichroic signal similar to that of previously observed in $\gamma$-$Fe_2O_3$, $BiFeO_3$ thin films, and $La_2FeCrO_6$ [60-62]. It has been shown that in $BiFeO_3$ weak ferromagnetism can be developed due to spin canting as a result of Dzyaloshinski-Moria interaction in the lattice. A similar XMCD spectrum has been reported by Kuo et. al., to describe the weak ferromagnetism in $BiFeO_3$ thin films due to canted antiferromagnetic ordering [61]. Gray *et al.* have also shown similar Fe $L_{2,3}$ XMCD signal in case of canted antiferromagnetic double perovskite $La_2FeCrO_6$ [62]. In case of $\gamma$-$Fe_2O_3$ the weak ferromagnetism arises as a result of alignment of spins of tetrahedral sites which is canted in nature [59, 63]. It is significant to mention here that in the *Pbnm* phase there are two sites available for Fe ions ($O_h$ and $T_d$) whereas in *R3c* phase only $O_h$ sites are available for Fe. Thus, the interfacial secondary *Pbnm* phase can also be responsible for the XMCD signal at Fe edge of 0.8BFO-0.2TMO at 180 K. The dilute antiferromagnetric (2D DAFF) layer can be understood to be due to the interaction between Fe ions in *Pbnm* phase which were observed from the IRM measurements. The canting of Fe spins close to the interface could arise due to the lattice strain at the interface as introduced by the TMO phase. This however should not affect the spin orientation further away from the interface. Gruber *et al.* have recently shown that the interfacial pinning plays a vital role in attaining the XMCD signal in Co/MnPc spinterface [64]. From the thickness dependent study of the layered structure it has been shown that the XMCD signal varies when the pinning layer varies [64]. From the XMCD spectra at 300 K and 180 K, it can be conferred that as we decrease the temperature from 300

K spin canting in the system increases in the system down to the spin reorientation temperature. Below the transition temperature the spins reorient in such a way that the weak ferromagnetism due to spin canting vanishes in the system. Thus, the XMCD signal can only be observed for 0.8BFO-0.2TMO composite at 180K as the spin reorientation transition temperature (T* = 169 K) is lower than 180 K whereas it is (T* = 215 K) higher than that of the XMCD measurement temperature (180 K) of 0.7BFO-0.3TMO.

Figure 8 (a-d) represents the Mn $L_{2,3}$-edge XAS and XMCD spectra of the two composites (0.7BFO-0.3TMO and 0.8BFO-0.2TMO) at 300 K and 180 K. The XAS spectra of the composites exhibit two broad spin-orbit split peaks of $L_3$ ($2P_{3/2}$) at ~ 641eV and $L_2$ ($2P_{1/2}$) at ~ 652eV separated by spin-orbit splitting energy ($\Delta E$ = ~ 11eV). The XAS spectra of the composites are compared with that of standard MnO sample from which it can be easily concluded that the $L_3$ and $L_2$ peaks are shifted to higher energy side for presence of higher valence state of Mn ions [65]. Interestingly, the composites show distinct behavior at $L_3$ and $L_2$ edges where the shape of the spectra does not match exactly with $Mn_2O_3$ in which Mn lie in 3+ state, rather it matches well with Mn XAS spectra observed in $La_{0.7}Sr_{0.3}MnO_3$ thin films in which Mn lie in a mix valence state [66]. The main peak observed at ~ 640 eV is due to the presence of $Mn^{3+}$ as the position matches well with Mn signal obtained from $Mn_2O_3$ in a tetragonally distorted $D_{4h}$ crystal field and from $TbMnO_3$ [67, 68]. The high intensity peak $2P_{3/2}$ observed at ~ 640eV shows distinct shoulder peaks at ~ 638 eV (marked by blue star) and 642 eV (marked by red arrow) indicating the existence of multivalent oxidation states of Mn. The spectral shape of any $L_{2,3}$ edge depends on different factors such as, local crystal field effects, multiplet structure given by the Mn 3d-3d and 2p-3d Coulomb and exchange interactions and the hybridization with the O 2p ligands [65]. The peaks can be assigned to presence of $Mn^{2+}$ and $Mn^{4+}$ respectively as has been done by several authors in Mn XAS spectra [66, 68-70]. To maintain the charge neutrality, affected by oxygen vacancy present in the lattice, the mixed oxidation states can evolve in the system. The co-valency can also arise from the charge transfer effect between the Mn 3d orbitals and O 2p ligand orbitals commonly observed in Mn based systems [57 and references therein]. The valence instability of $Mn^{3+}$ ions can also give rise to creation of $Mn^{4+}$ and $Mn^{2+}$ species which is known to modify the bulk magnetic and electrical properties of different manganite systems [66]. Oxygen vacancies present in the composites which is observed in XPS analysis (not shown here) of the composite can change the effective superexchange (SE) interaction between Mn ions. Yang *et al*. reported the change in magnetic spin structure of $TbMnO_3$ due to the alternation in SE between next-nearest-neighbor Mn ions mediated through O ion [71, 72].

The XAS spectra of the 0.8BFO-0.2TMO (figure 8 b and d) show more pronounced shoulder peaks due to the presence of $Mn^{2+}$ and $Mn^{4+}$ ions (marked by blue star and red arrow). From the spectra it can be concluded that in this composite the 2+ state of Mn ion dominated over other valence states. Moreover, the intensity ratio of the $L_3$ and $L_2$ (i.e. $L_3/L_2$) edges is also a significant parameter to determine the dominating oxidation states of transition metal and their oxides (TMs) with 3d occupancy [73]. For the transition metals which have $d^0$ to $d^5$ occupancy (i.e., up to half filled occupancy) the increment in $L_3/L_2$ intensity ratio signifies reduction in oxidation state of the TMs [73, 74]. Larger $L_3/L_2$ intensity ratio for Mn ion in 0.8BFO-0.2TMO composite than the other reveals that the dominating state of Mn is $Mn^{+2}$. The presence of large amount of $Mn^{2+}$ and $Mn^{4+}$ in this composite is understood to be due to the large lattice strain experienced by the $TbMnO_3$ due to higher concentration of BFO in the 0.8BFO-0.2TMO composite. Lattice strain is known to create oxygen vacancies in a lattice which in turn creates the mix valence state of Mn ion in the composites [75, 76].

Mn $L_{3,2}$ edge XMCD signals of the composites are shown in the bottom panel of figure 8 (a-d). The room temperature XMCD signal have been multiplied by 5 for better visualization which show very weak XMCD signal at $L_3$ and $L_2$ edge. Although the XMCD signals are weak and noisy at room temperature XMCD peaks can be found in opposite polarity for $Mn^{2+}$ with respect to that of $Mn^{3+}$ and $Mn^{4+}$ for both the composite which show the opposite alignment of $Mn^{2+}$ ionswith respect to $Mn^{3+}$ and $Mn^{4+}$. Interestingly, the 180 K XMCD spectrum of 0.8BFO-0.2TMO shows prominent signal just like its Fe counterpart. The signal for $Mn^{2+}$ is opposite in nature to the signal of $Mn^{3+}$ and $Mn^{4+}$ showing the antiferromagnetic coupling between them. Further, it is found that the magnitude of the Mn dichroism is almost diminished at 180K for the composite 0.7BFO-0.3TMO due to the spin reorientation transition occurring at a higher temperature (225 K). This reduction of the Mn dichroism could be understood by taking into account spin-reorientation transition near ~200K observed from the magnetization measurement. Interestingly, the 180 K XAS signal of 0.7BFO-0.3TMO shows increase in the intensity of the shoulder peak at ~638 eV which signifies the increase in the charge transfer process between Mn ions at low temperature.

The contribution of spin moments and the orbital moments to XMCD signal are calculated following the XMCD sum rule [77]. It is found that the contribution of orbital momentum is not negligiblein comparison to the spin momentum in all the cases. This large orbital momentum can break the local symmetry in the proximity of the BFO/TMO interface, leading to strongly enhanced unidirectional anisotropy energy [78]. This local anisotropy energy is strong enough to induce effective exchange imbalance at the interface, which try to rotate spin moments by spin-orbit- coupling (SOC) [79]. Recently, Nistor et al. have reported that the exchange bias in their system arises from the coupling of the Mn spins to the uncompensated spin at the interface [79]. Thus, in our sample also there might be coupling of Mn spins to the uncompensated Fe spins at the interface. In addition to the coupling of Mn and Fe spins there are also exchange coupling between Mn-Mn ions. This coupling of Mn-Mn and Mn-Fe ions would give rise to layer of pinned magnetic moments at the interface. However, the detection and measurement of pinned magnetic moments cannot be done following the simple XMCD measurement protocol [78].

### D. UV-Vis Spectroscopy:

The presence of different charge state of TM ions at the interface due to the charge transfer between Mn-Mn ions may lead to the band reconstruction at the interface. As a result of the reconstruction band gap of the composite is expected to decrease [80, 81]. Thus to confirm the band gap reduction we have measured the absorption spectrum in the UV and visible range. For studying the absorption characteristics of the composite, absorbance at different wavelength (k) (range of 200–800 nm) were recorded and the absorption coefficients (a) were calculated at corresponding wavelengths. As can be seen from the spectrum the absorption band edge lies beyond the range of measurement. We have followed the Tauc's method to estimate the band gap of the composite from the absorption spectra [82]. The photon energy (hν) and the band gap energy for a particular transition are related by the equation;

$$(\alpha h\nu) = K(h\nu - E_g)^n \qquad (3)$$

where α is the absorption coefficient given by α = 2.303 (Ab/t), (here, Ab is absorbance and t is thickness of the cuvette which is 1 cm), K is the edge width parameter. The value of n depends on the type of transition, i.e., allowed direct, allowed indirect, forbidden direct, and forbidden indirect for which it can have values 1/2, 2, 3/2, and 3, respectively. $BiFeO_3$ is known to have a bang gap of 2.1 eV to 2.7 eV in different form of the material such as bulk, nanomaterials, or single crystalline material [83-85]. Since $BiFeO_3$ and $TbMnO_3$ both are known to be direct band gap material, the band gap of the composite was determined from the linear fitting of the straight line part of the $(\alpha h\nu)^2$ versus photon energy (hν) plot on the hν axis [86]. From the Tauc's plot (figure 9) it is evident that the band gap of the material lies in the rage of ~ 0.9-1.0 eV which is very low compared to other reported value for $BiFeO_3$ [83-85]. Therefore the reduction in the band gap of the composite system confirms the band reconstruction phenomenon due to the charge transfer between the TM ions in the composite.

Thus, analyzing the UV-Vis absorption spectrum and XAS, XMCD spectra of all the $L_{3,2}$ edges, it can be concluded that there is charge transfer between Mn ions due to oxygen vacancy and instability of the $Mn^{3+}$ oxidation state. The charge transfer results in a mix valence state of Mn ions in the composite. As a result of the transfer and mix valence state of the TM ions band gets reconstructed at the interface and the band gap of the material drops. The presence of mix valence state creates different superexchange interaction between TM ions across the interface and in bulk. Interestingly, our M-H data measured following N-type and P-type measurement protocol (inset of figure 2) shows the existence of pinned magnetic moments which is not rotatable on applying different magnetic fields. The result indicates the coupling of the Mn spins to the uncompensated spin of Fe which forms a pinned layer of spin moments at the interface of the two materials at 300 K. This weak ferromagnetic layer is stabilized against the thermal fluctuations through exchange coupling to the 2D DAFF layer and/or to the core canted antiferromagnetic spins of BFO. On lowering the temperature from 300 K the spin canting increases as a result the uniaxial antiferromagnetic anisotropy decreases (seen from XMCD results). The pinned ferromagnetic layer on the other hand gets stronger on decreasing the temperature as the thermal fluctuation on the spins decreases. The exchange bias of the system increases up to T* on increasing the temperature with stronger exchange coupling as the AFM anisotropy increases in that temperature range. Further increase in the temperature results in decrease in the exchange bias due to the spin reorientation transition. The pinned layer getting weaker due to thermal fluctuations also affects the value of EB. The temperature at which the maximum $H_{SEB}$ observed is lower than 0.7BFO-0.3TMO in the composite 0.8BFO-0.2TMO since spin reorientation is found to be at a lower temperature. It has also smaller amount of TMO present in it which creates thinner layer of pinned moments. Thus, the exchange bias is developed as a result of pinning of the interfacial ferromagnetic spins to the core antiferromagnetic BFO spins. Due to the dual phase structure of BFO a thin layer of DAFF is formed at the skin of BFO core which also favours the formation of canted antiferromagnetic ordering of the BFO lattice. Moreover, the role of interfacial spins has also been confirmed by the nature of training effect of the exchange bias.

## 4. Conclusion

In summary, we have studied the origin of exchange bias induced in $BiFeO_3$-$TbMnO_3$ composite (70:30 and 80:20 stoichiometric ratio) prepared via solid state reaction. In the prepared composite $BiFeO_3$ has the rhombohedrally distorted perovskite (*R3c*) structure at the core and orthorhombically distorted *Pbnm* structure at the shell or skin whereas $TbMnO_3$ has orthorhombically distorted perovskite (*Pbnm*) structure. We observed large spontaneous exchange bias at room temperature which could not be explained on the basis of the pinning or a super-spin-glass state at low temperatures. Different models explaining the SEB have been investigated, such as core-shell of hard and dilute antiferromagnet and charge transfer between Fe and Mn ions present at the interface of the two materials in the composite. The charge transfer between the Mn-Mn ions at the interface takes place due the oxygen vacancy and instability of $Mn^{3+}$ ions. The presence of mix charge state creates different exchange interaction between the TM ions (ferro and antiferromagnetic). The strong magnetoelectric coupling between the two materials initiates canting and pinning of the interfacial BFO or TMO spins. From the IRM measurement signatures of the presence of 2D dilute antiferromagnet has been found which plays a significant role in obtaining the exchange bias. The 2D DAFF layers are the result of the formation of interfacial phase of BFO (*Pbnm*) in which the antiferromagnetic ordering is different than that of the *R3c* phase in the bulk as result the Fe edge shows $\gamma$-$Fe_2O_3$ like XMCD signal. The large SEB obtained at different temperatures also shows nonmonotonic variation with temperature. The reduction in $H_{SEB}$ values below maximum $H_{SEB}$ obtained at 225

K and 180 K has been understood as the reduction in interfacial coupling below the spin reorientation temperature observed at ~ 215 K and 169 K from the M-T measurement in ZFC an FC mode as well as from the XMCD measurements at 180 K. The decrease in AFM anisotropy due to spin canting and increase in FM ordering in the pinned layer with decrease in the temperature have been found to influence the temperature dependence of the EB. The role of interfacial pinning layer which influences the exchange bias in the system has also been confirmed by the nature of M-H loops measured via different protocols, the training effect of the exchange bias and XMCD spectra of Fe and Mn $L_{2,3}$ edge. Moreover, the results from training effect measurement show stable value of exchange bias after several loops of hysteresis measurement which is a good quality for application in different devices.

## Acknowledgement

Authors would like to acknowledge the central instrument facility centre of IIT (BHU) for the magnetic measurements.

**Figure Captions:**

**Figure 1**: Rietveld refinement of the XRD data of the composite 0.7BFO-0.3TMO.

**Figure 2**: The hysteresis loop shift of 0.7BFO-0.3TMO composite measured at different temperature across 5-300K after zero-field cooling. The inset shows M-H loop recorded at 300 K following two measurement protocols (P-type and N-type)

**Figure 3**: The temperature dependence of exchange bias field ($H_{SEB}$) and coercivity and ($H_C$) for **(a)** 0.7BFO-0.3TMO **(b)** 0.8BFO-0.2TMO. Inset shows the close view of the M-H loops of 0.7BFO-0.3TMO at (a) 300 K, (b) 180 K and 0.8BFO-0.3TMO at (c) 300 K and (d) 180 K

**Figure 4:** ZFC and FC magnetization vs. temperature plots for **(a)** 0.7BFO-0.3TMO and **(b)** 0.8BFO-0.2TMO composite under an applied magnetic field of 1000 Oe. Inset shows the dM/dT vs T plots for the respective sample. The arrows in the insets point towards the spin reorientation transition temperature T*.

**Figure 5**: Training effect of SEB at 300 K of (a) 0.7BFO-0.3TMO and (b) 0.8BFO-0.2TMO. Inset shows $H_{SEB}$ vs. n for the respective sample where solid red lines show the fitting of the experimental data to Eq. 1 and blue line to Eq. 2.

**Figure 6:** Isothermal remnant magnetization (IRM) and the virgin loop of magnetization vs. magnetic field of the composite 0.7BFO-0.3TMO.

**Figure 7:** Fe $L_{2,3}$ XAS and XMCD spectra of **(a)** 0.7BFO-0.3TMO at 300 K **(b)** 0.8BFO-0.2TMO at 300 K **(c)** 0.7BFO-0.3TMO at 180 K **(d)** 0.8BFO-0.2TMO at 180 K. XMCD signal was multiplied by 5 in all the cases for better visualization. Inset in **(a)** and **(b)** shows the comparison of $t_{2g}$ peak of $L_3$ edge and $L_2$ edge of the composites with standard $\alpha$-$Fe_2O_3$ respectively.

**Figure 8**: Mn $L_{2,3}$ XAS and XMCD spectra of **(a)** 0.7BFO-0.3TMO at 300 K **(b)** 0.8BFO-0.2TMO at 300 K **(c)** 0.7BFO-0.3TMO at 180 K **(d)** 0.8BFO-0.2TMO at 180 K. XMCD signal was multiplied by 5 in case of **(a)** and **(b)** for better visualization. Inset in **(a)** shows the comparison between XAS spectra from 0.7BFO-0.3TMO, 0.8BFO-0.2TMO and standard α-$Fe_2O_3$ sample.

**Figure 9**: Tauc plot for the determination of the optical band gap of the composite 0.7BFO-0.3TMO at room temperature. The black dashed arrow is guide to the eye, showing the extracted bad gap. The inset shows the absorption spectrum of the composite from which the Tauc plot has been estimated.

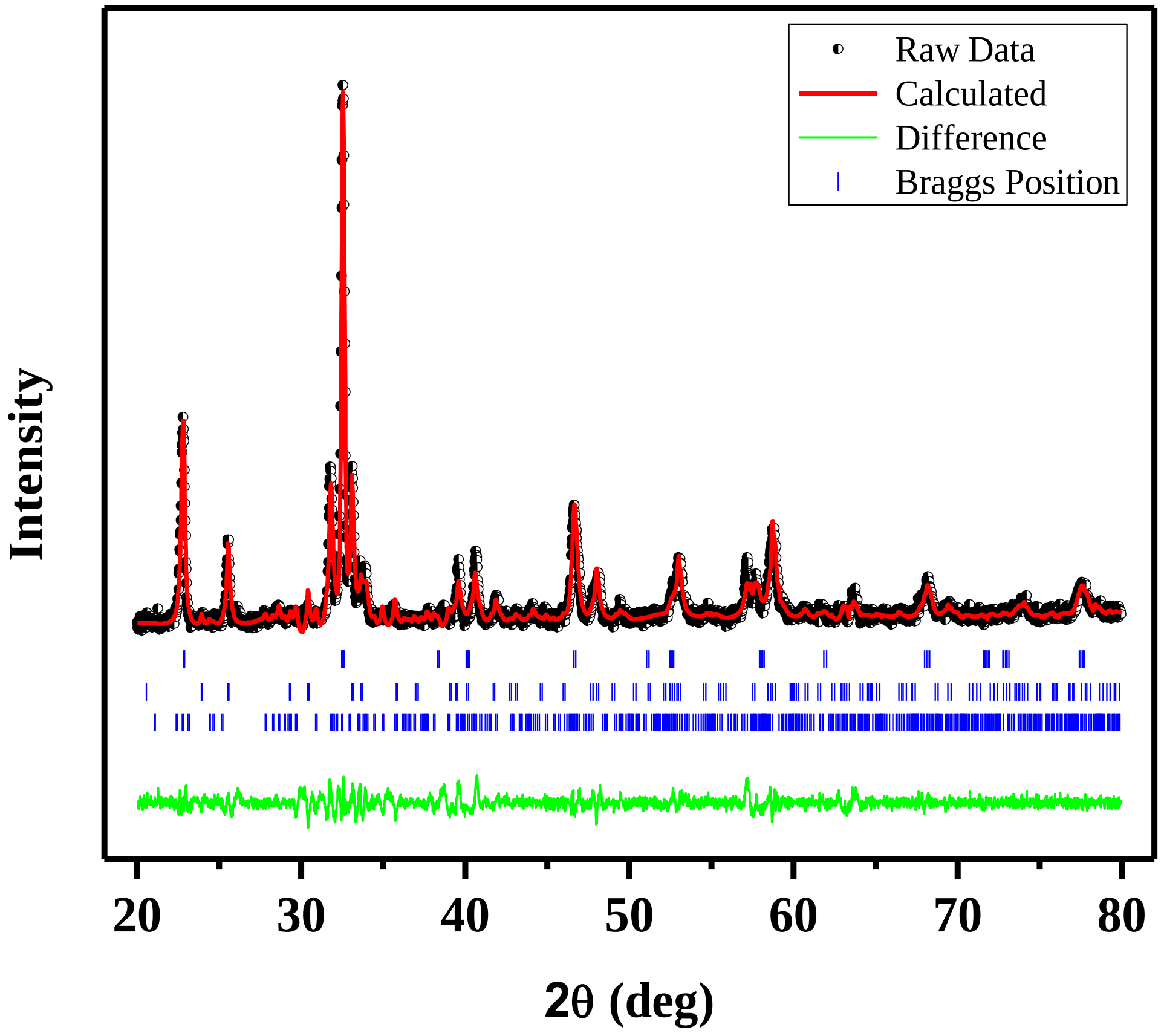

**Figure. 1**: Rietveld refinement of the XRD data of the composite 0.7BFO-0.3TMO.

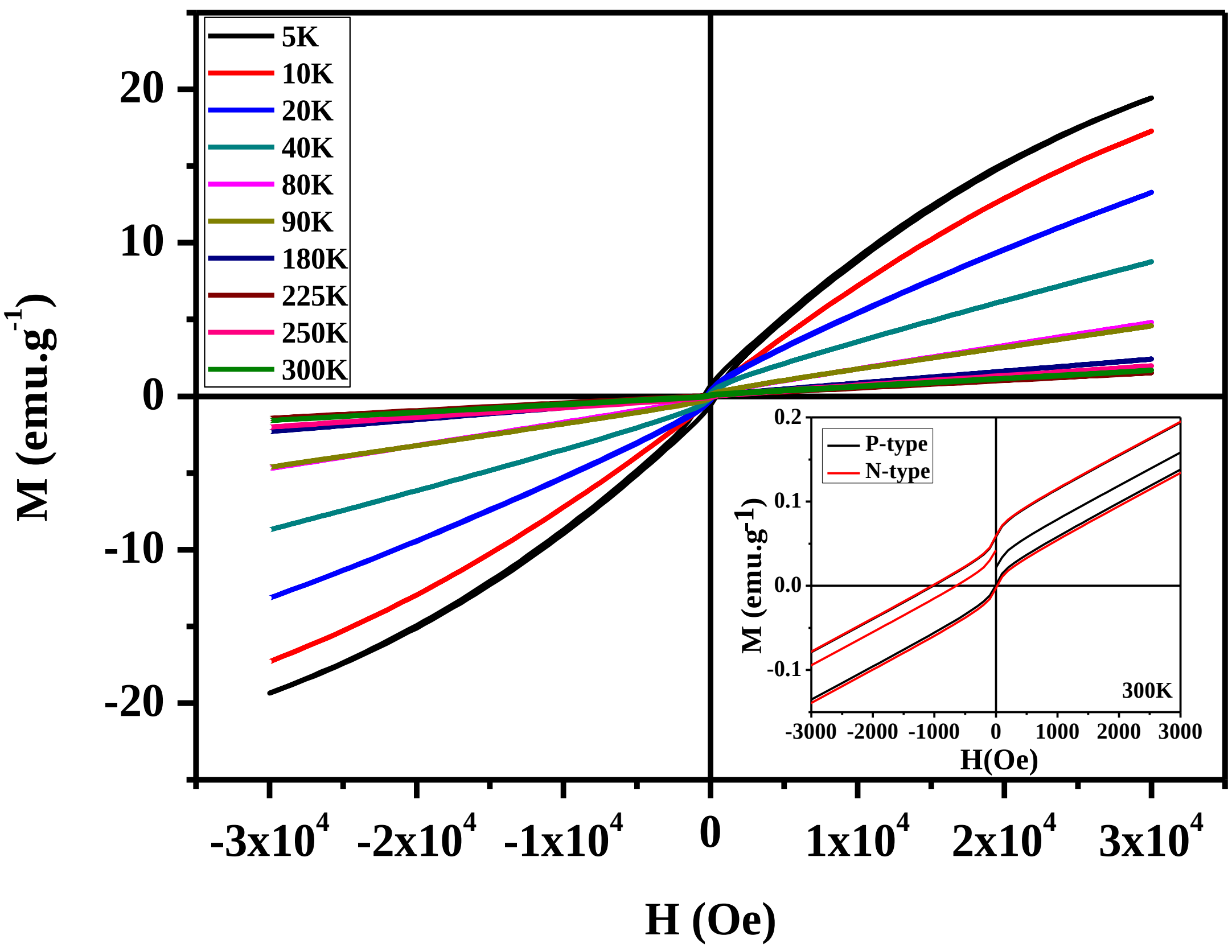

**Figure 2**: The hysteresis loop shift of 0.7BFO-0.3TMO composite measured at different temperature across 5-300K after zero-field cooling. The inset shows M-H loop recorded at 300 K following two measurement protocols (P-type and N-type)

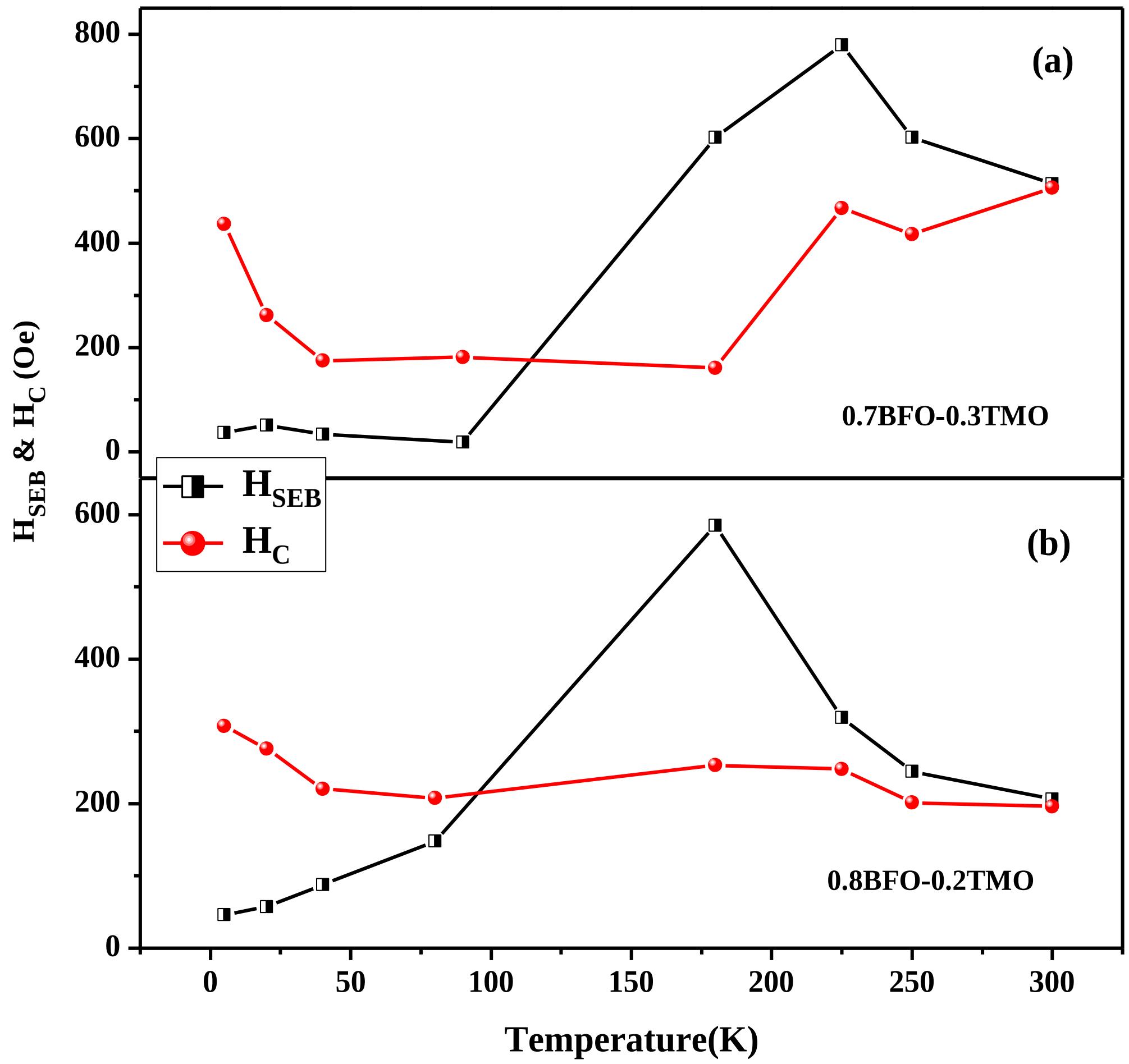

**Figure 3**: The temperature dependence of exchange bias field ($H_{SEB}$) and coercivity of (a) 0.7BFO-0.3TMO and (b) 0.8BFO-0.2TMO.

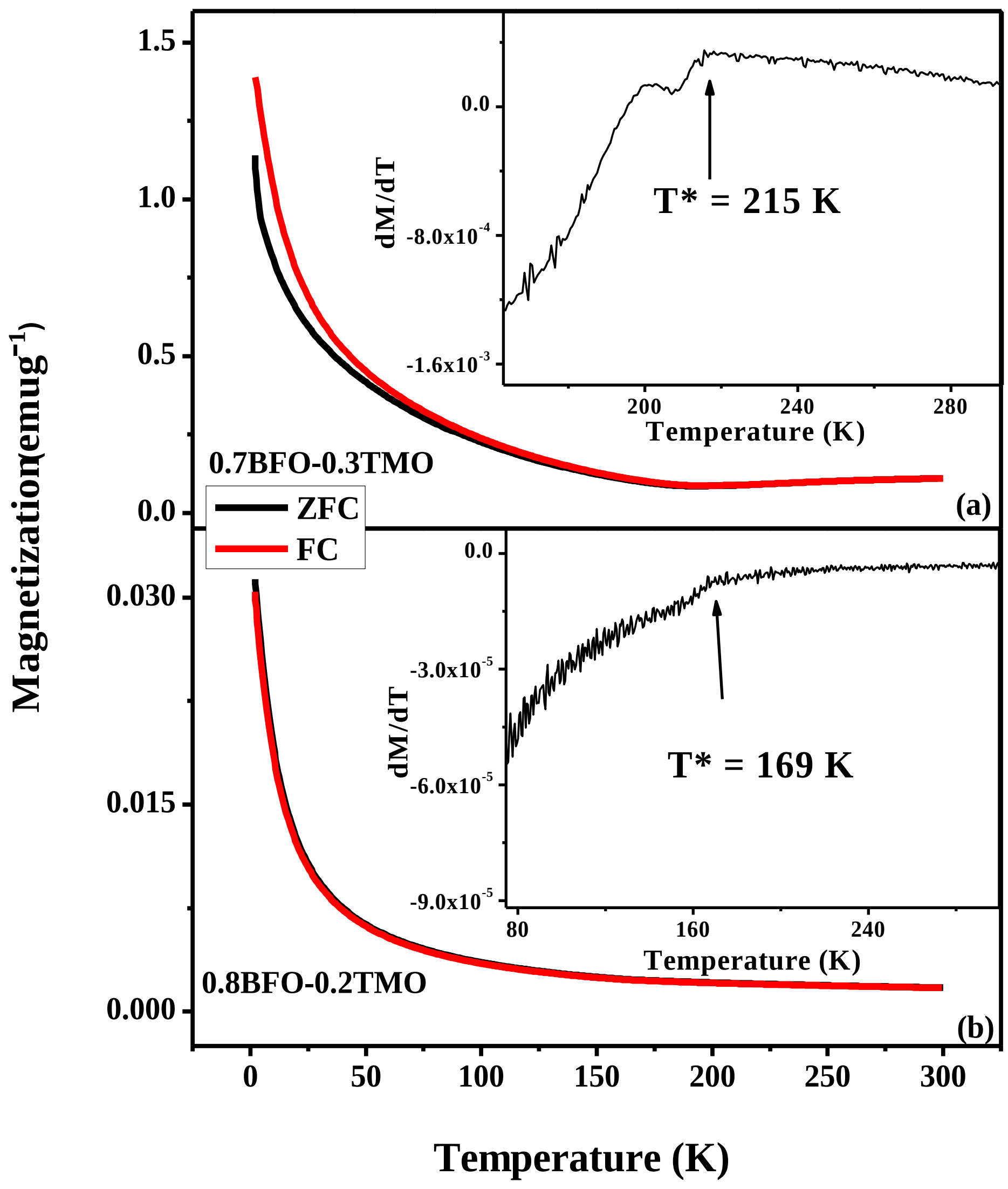

**Figure 4**: ZFC and FC magnetization vs. temperature plots for **(a)** 0.7BFO-0.3TMO and **(b)** 0.8BFO-0.2TMO composite under an applied magnetic field of 1000 Oe. The insets in **(a)** and **(b)** show the spin reorientation transition for both the composites at 215 K and 169 K respectively.

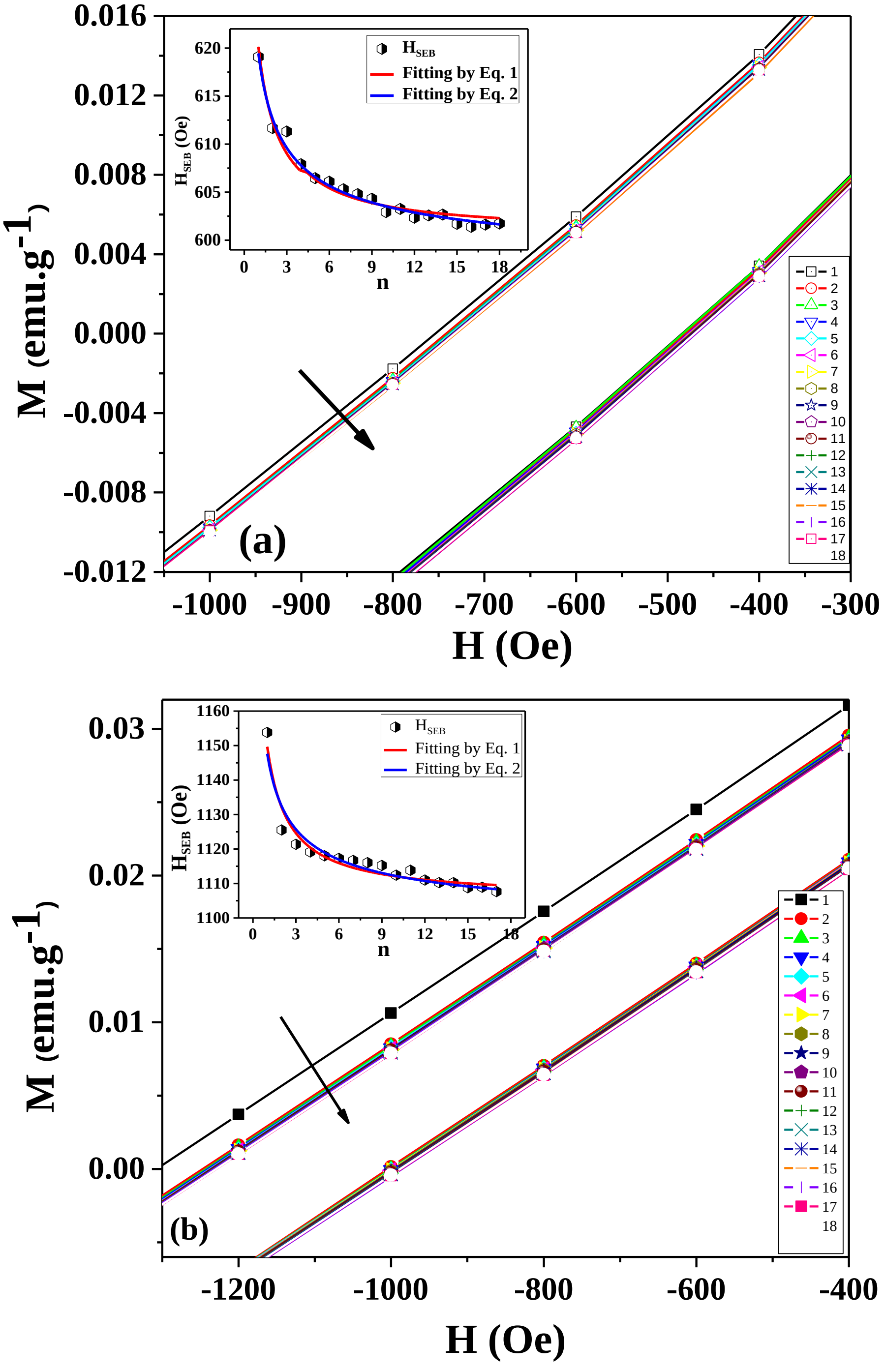

**Figure 5:** Training effect of SEB at 300 K of (a) 0.7BFO-0.3TMO and (b) 0.8BFO-0.2TMO. Inset shows $H_{SEB}$ vs. n for the respective sample where solid red lines show the fitting of the experimental data to Eq. 1 and blue line to Eq. 2.

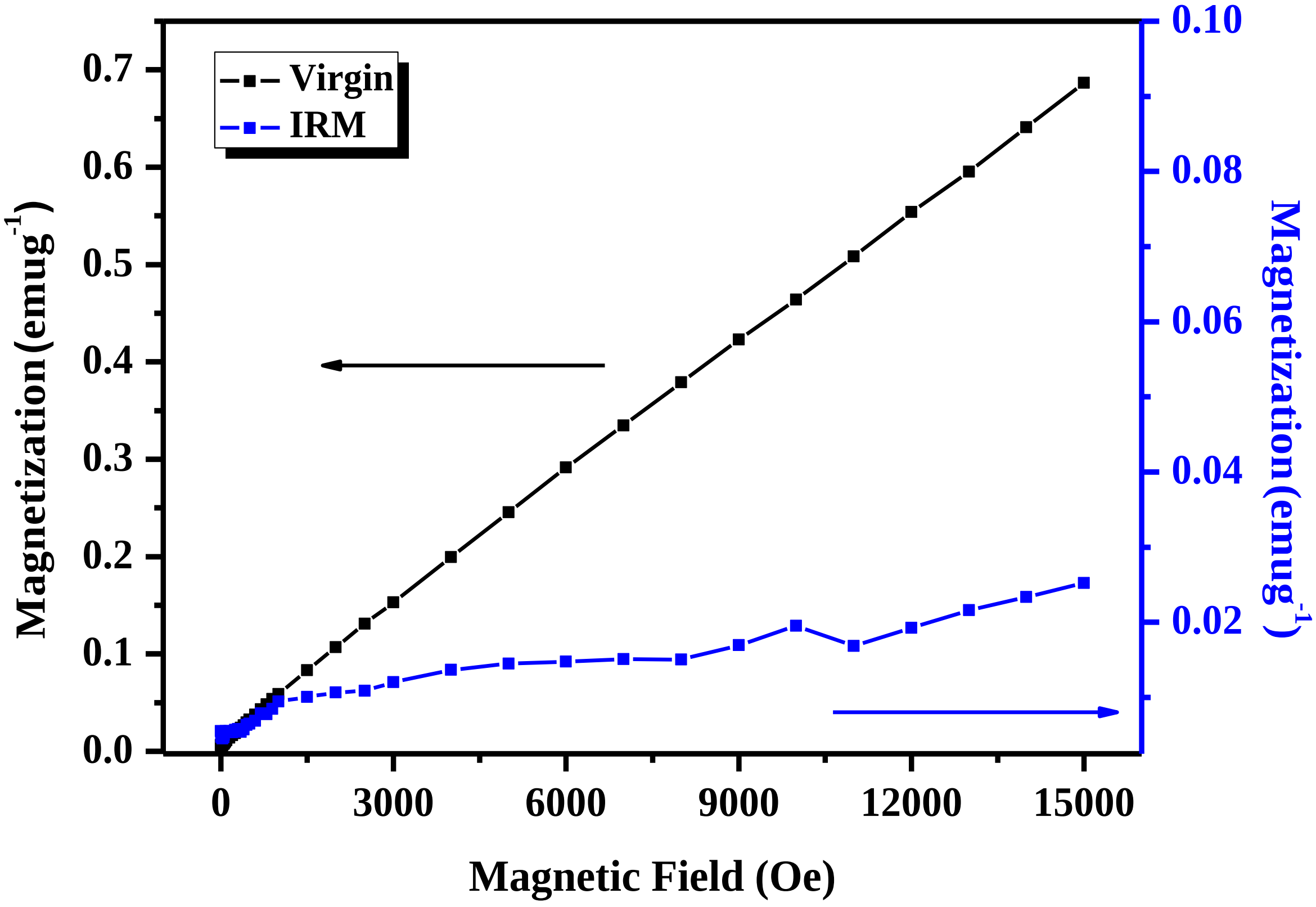

**Figure 6**: Isothermal remnant magnetization (IRM) and the virgin loop of magnetization vs. magnetic field of the composite 0.7BFO-0.3TMO.

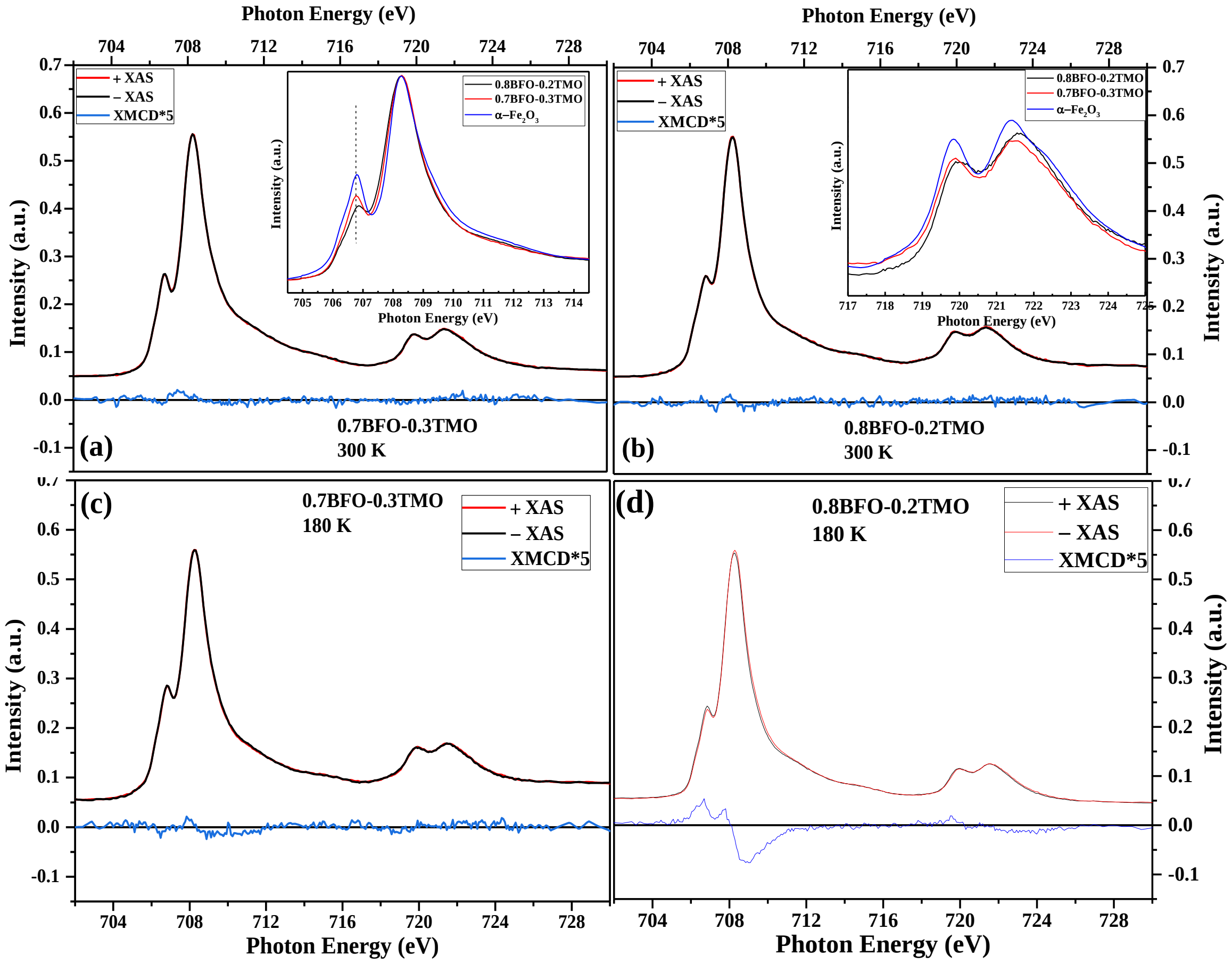

**Figure 7**: Fe $L_{2,3}$ XAS and XMCD spectra of **(a)** 0.7BFO-0.3TMO at 300 K **(b)** 0.8BFO-0.2TMO at 300 K **(c)** 0.7BFO-0.3TMO at 180 K **(d)** 0.8BFO-0.2TMO at 180 K. XMCD signal was multiplied by 5 in all the cases for better visualization. Inset in **(a)** and **(b)** shows the comparison of $t_{2g}$ peak of $L_3$ edge and $L_2$ edge of the composites with standard α-$Fe_2O_3$ respectively.

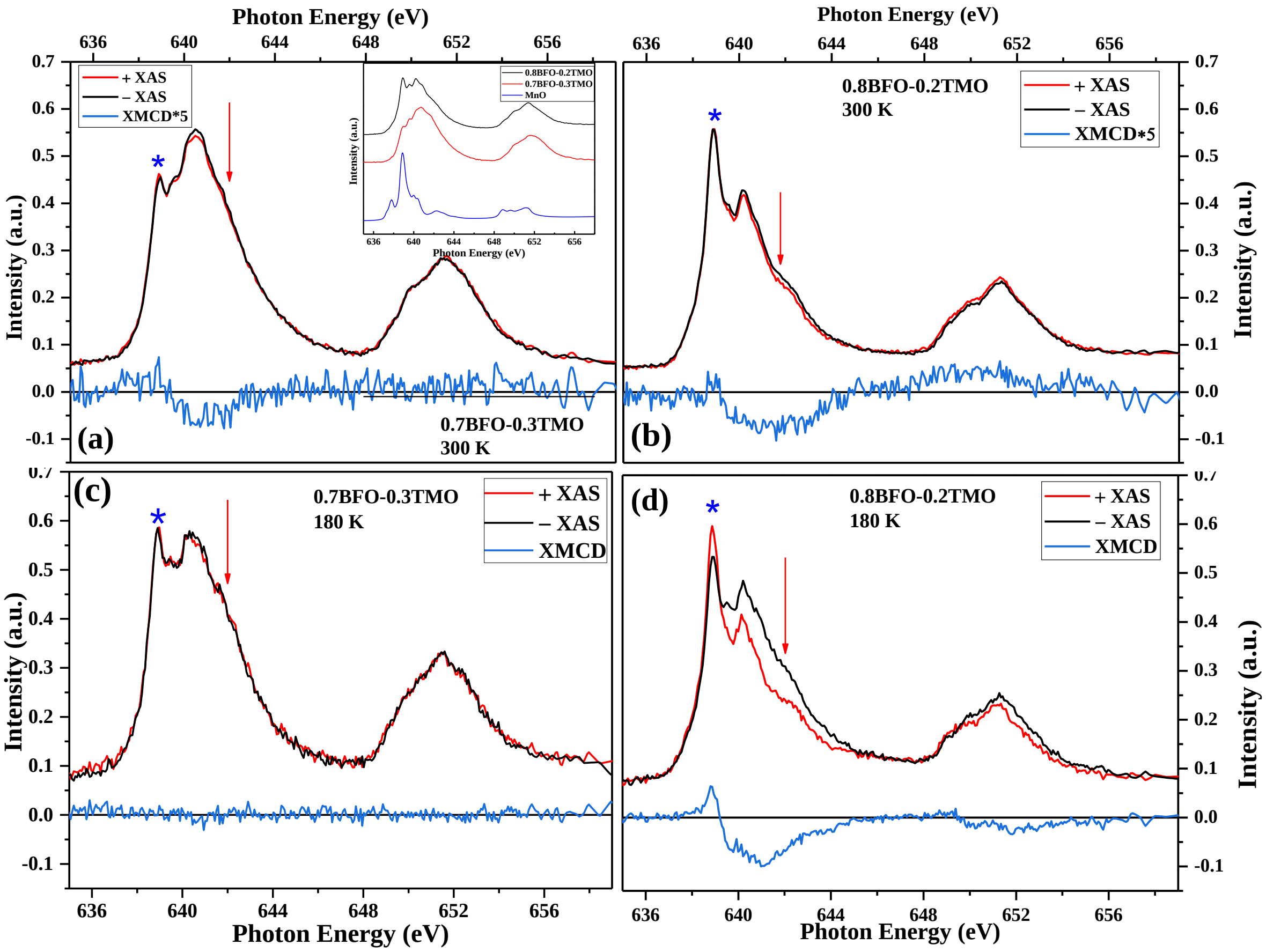

**Figure 8**: Mn $L_{2,3}$ XAS and XMCD spectra of **(a)** 0.7BFO-0.3TMO at 300 K **(b)** 0.8BFO-0.2TMO at 300 K **(c)** 0.7BFO-0.3TMO at 180 K **(d)** 0.8BFO-0.2TMO at 180 K. XMCD signal was multiplied by 5 in case of **(a)** and **(b)** for better visualization. Inset in **(a)** shows the comparison between XAS spectra from 0.7BFO-0.3TMO, 0.8BFO-0.2TMO and standard α-$Fe_2O_3$ sample.

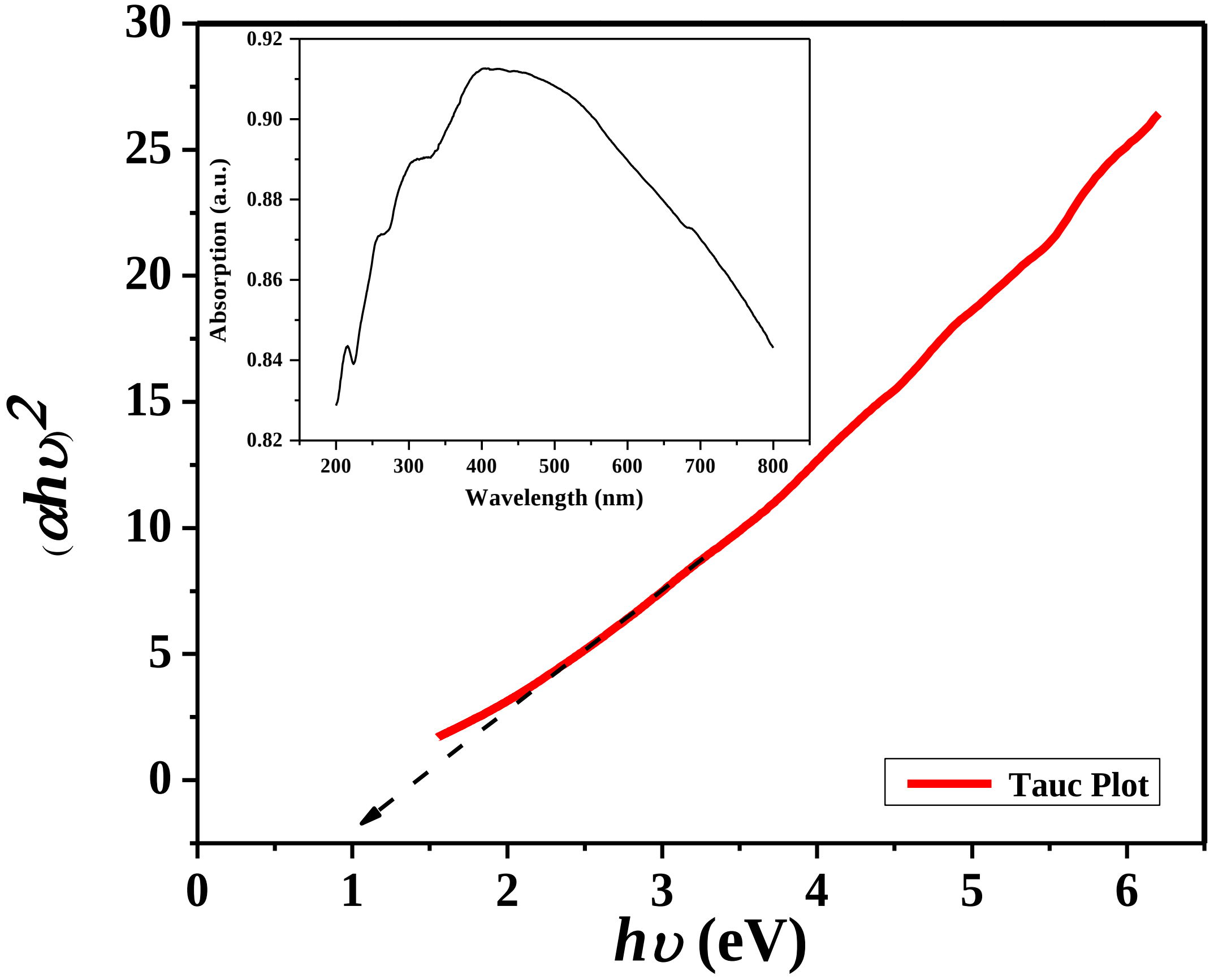

**Figure 9**: Tauc plot for the determination of the optical band gap of the composite 0.7BFO-0.3TMO at room temperature. The black dashed arrow is guide to the eye, showing the extracted bad gap. The inset shows the absorption spectrum of the composite from which the Tauc plot has been estimated.